# Spatio-Temporal Dual-Stream Neural Network for Sequential Whole-Body PET Segmentation

Kai-Chieh Liang[a], Lei Bi[a, b, *], Ashnil Kumar[b, d], Michael Fulham[a, b, c], Jinman Kim[a, b]

[a] School of Computer Science, University of Sydney, NSW, Australia

[b] Australian Research Council Training Centre for Innovative BioEngineering, NSW, Australia

[c] Department of Molecular Imaging, Royal Prince Alfred Hospital, NSW, Australia

[d] School of Biomedical Engineering, University of Sydney, NSW, Australia

[*] Corresponding author: lei.bi@sydney.edu.au

*Abstract*— Sequential whole-body 18F-Fluorodeoxyglucose (FDG) positron emission tomography (PET) scans are regarded as the imaging modality of choice for the assessment of treatment response in the lymphomas because they detect treatment response when there may not be changes on anatomical imaging. Any computerized analysis of lymphomas in whole-body PET requires automatic segmentation of the studies so that sites of disease can be quantitatively monitored over time. State-of-the-art PET image segmentation methods are based on convolutional neural networks (CNNs) given their ability to leverage annotated datasets to derive high-level features about the disease process. Such methods, however, focus on PET images from a single time-point and discard information from other scans or are targeted towards specific organs and cannot cater for the multiple structures in whole-body PET images. In this study, we propose a spatio-temporal 'dual-stream' neural network (ST-DSNN) to segment sequential whole-body PET scans. Our ST-DSNN learns and accumulates image features from the PET images done over time. The accumulated image features are used to enhance the organs / structures that are consistent over time to allow easier identification of sites of active lymphoma. Our results show that our method outperforms the state-of-the-art PET image segmentation methods.

*Index Terms*— spatio-temporal, medical image segmentation, positron emission tomography (PET)

## 1. INTRODUCTION

The lymphomas are a heterogeneous group of tumors of the lymphatic system. They usually involve lymph nodes but can involve the spleen, the bone marrow and any other organ in the body. They can be relatively slow-growing or rapidly progressive and are glucose-avid on positron emission tomography (PET) and so are easy to identify, in most cases, on PET scans [1]. $^{18}$F-Fluorodeoxyglucose (FDG) PET is regarded as the imaging modality of choice to determine the extent of disease (stage) of the lymphomas [1, 2]. Further, given the ability of PET to detect disease in structures that are not enlarged, it is the preferred modality to assess response to therapy over sequential PET scans [3]. The interpretation of sequential PET can be problematic given that in some structures there is marked physiological FDG uptake (brain and heart) and FDG is excreted via the urinary tract and there is FDG excretion noted in the kidneys and ureters and there is pooling of FDG in the bladder. As such, sites of FDG excretion and physiologic uptake, that we denote as sFEPU, could potentially be confused as sites consistent with high grade tumors [4]. Computerized algorithms that can detect, segment, and then remove the sFEPU will improve the interpretation of PET scans. However, this is a difficult task as sFEPU exhibit subtle variations in FDG uptake across different patients, and the aforementioned similarity of sFEPU uptake to tumor uptake carries the risk that tumors may be inadvertently removed.

We hypothesis that sequential scans can be leveraged to address these difficulties, using machine learning to model the spatio-temporal consistencies of sFEPU, where the sFEPU will not be markedly affected by treatment; whereas sites of abnormal uptake due to disease are more likely to change [5]. We suggest that these characteristics can be used to improve sFEPU segmentation.

### 1.1 Related Work

The sFEPU segmentation relates to prior works on PET image segmentation. Traditional PET image segmentation methods focus on using thresholding, graph based models e.g., graph cut, and various classifiers e.g., support vector machine (SVM) [6-10]. Vauclin et al. compared three thresholding methods based on standardized uptake value (SUV) and contrast features for segmenting tumor volumes from PET images [6]. Drever et al. found that thresholding based methods have better segmentation performance compared with Sobel edge and watershed based segmentation methods with a PET phantom study [7]. Thresholding based methods, however, are limited by the distribution of the regions of interest (ROIs) and may fail if the distribution contains multiple peaks (e.g., inhomogeneous ROIs). Ju et al. proposed to use graph cut algorithm separately on PET and computed tomography (CT) images and the segmentation results derived from these two images were then ensembled [8]. Specifically, user defined seeds were selected from PET and CT images for initialization and the seeds were then grown or morphed to the tumor boundaries according to the graph cut algorithm. However, such manual initializations are usually subjective, time-consuming, and non-reproducible. As a consequence, such methods are unreliable for wide adoption in clinical environments. Gubbi et al. used SVM to detect and localize potential tumors from PET images and the boundaries of detected tumors were then refined with deformable model [9]. However, the performance of the classifier based methods is heavily reliant on correctly tuning a large number of parameters and effective pre-processing techniques such as image denoising and smoothing, which thereby restricting its generalizability.

State-of-the-art PET image segmentation is based on convolutional neural networks (CNNs) [11, 12]. This success is attributed to the ability of the CNNs to leverage large datasets to derive image feature representations that have high-level semantics [13-15]. There have been several CNN methods for PET image segmentation. Blanc-Durand et al. used a 3D U-Net to segment gliomas in brain PET images [16]. Bi et al. used pre-trained CNN on natural images for segmenting sFEPU regions on lymphoma PET images [17]. Taghanaki et al. proposed a combo loss (combining cross-entropy loss and dice loss) with a 3D CNN to segment multiple organs such as the brain, heart, left kidney, right kidney, and bladder in head-and-neck cancer PET images [18]. Haigen et al. proposed a 3D multi-view CNN to segment lymphomas  [19]. All these methods, however, were designed for a single time-point. In our literature search, we did not identify any PET segmentation methods designed for sequential PET data but did identify methods for other imaging modalities. Debus et al. used a 3D CNN to classify cardiac phases and to segment the wall of the left ventricle from sequential magnetic resonance (MR) images, where the image features from different time-points were concatenated to jointly learn spatio-temporal information [20]. Zhang et al. used a convolutional long short term memory (ConvLSTM) architecture for tumor growth prediction on sequential CT scans [21]. These methods, however, focused on modeling a particular organ or structure (e.g., heart and tumors) and did not cater for multiple organs or structures. In addition, these methods were not designed for PET images which exhibit largely different imaging characteristics such as comparatively lower signal-to-noise ratios and image resolutions.

## *1.2 Our Contributions*

We propose a spatio-temporal dual-stream neural network (ST-DSNN) method for the automated sFEPU segmentation, with the following contributions to the state-of-the-art:

- an architecture with two CNNs that process each time-point independently in its own stream and then fuse the streams and hence learns the pixel-to-pixel correlation between the two images in the sequence and,

- modelling spatio-temporal consistency information from sequential PET images to improve segmentation accuracy.

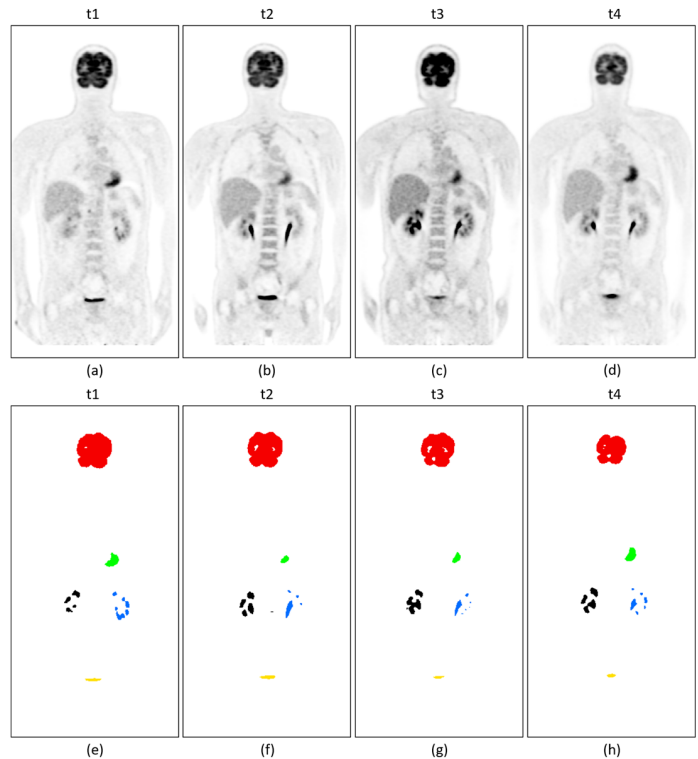

Figure 1. Sequential PET images in a patient with lymphoma. The top row (a)-(d) show PET images at time-points t1 to t4, and the bottom row (e)-(h) are the corresponding sFEPU annotations with red for brain uptake, green for heart, blue for left kidney, black for right kidney and yellow for bladder.

## 2. Methods

### 2.1 Materials

Our dataset consists of sequential PET images from 10 lymphoma patients: 6 patients with 2 scans; 3 patients with 3 scans and 1 patient with 4 scans. The small dataset is mainly attributed to the complexity of the sequential data acquisition process and the challenges in sequential data annotation process. Although the dataset is small, we suggest that it is appropriate to demonstrate the effectiveness of the proposed method. From this dataset, we obtained 21 unique PET image pairs for training and testing, where training and testing datasets were separated by patients. The image pairs are shown in Table 1. The dataset was acquired from the Department of Molecular Imaging, Royal Prince Alfred (RPA) Hospital, Sydney, NSW, Australia. All the images were acquired on a Biograph True-Point PET-CT scanner (Siemens Medical Solutions, Hoffman Estates, IL, USA). PET images were reconstructed using the 3-D ordered-subset expectation-maximization (3-d-OSEM) method. The reconstructed PET image had a resolution of $168 \times 168$ pixels with a pixel size of $4.07$ mm$^2$ and the slice thickness of 3 mm. The CT images were not used in this study. All data were de-identified. sFEPU were manually annotated by a researcher based on clinical reports and example annotations are shown in Fig. 1 where different colors are used to display the different organs. We aligned all the images for each patient center-to-center by manually removing the first and the last few slices such that all patient PET studies had the same dimensions and so all the images could be paired across the sequential scans. We did not remove images slices that contained sFEPUs.

Table 1. Image pairs for our ST-DSNN.

|  | Stream 1 | Stream 2 |
|---|---|---|
| **2 scans (6 patients)** | Scan 1 | Scan 2 |
| **3 scans (3 patients)** | Scan 1 | Scan 2 |
|  | Scan 1 | Scan 3 |
|  | Scan 2 | Scan 3 |
| **4 scans (1 patient)** | Scan 1 | Scan 2 |
|  | Scan 1 | Scan 3 |
|  | Scan 1 | Scan 4 |
|  | Scan 2 | Scan 3 |
|  | Scan 2 | Scan 4 |
|  | Scan 3 | Scan 4 |

## 2.2  Network architecture of spatio-temporal dual-stream neural network (ST-DSNN)

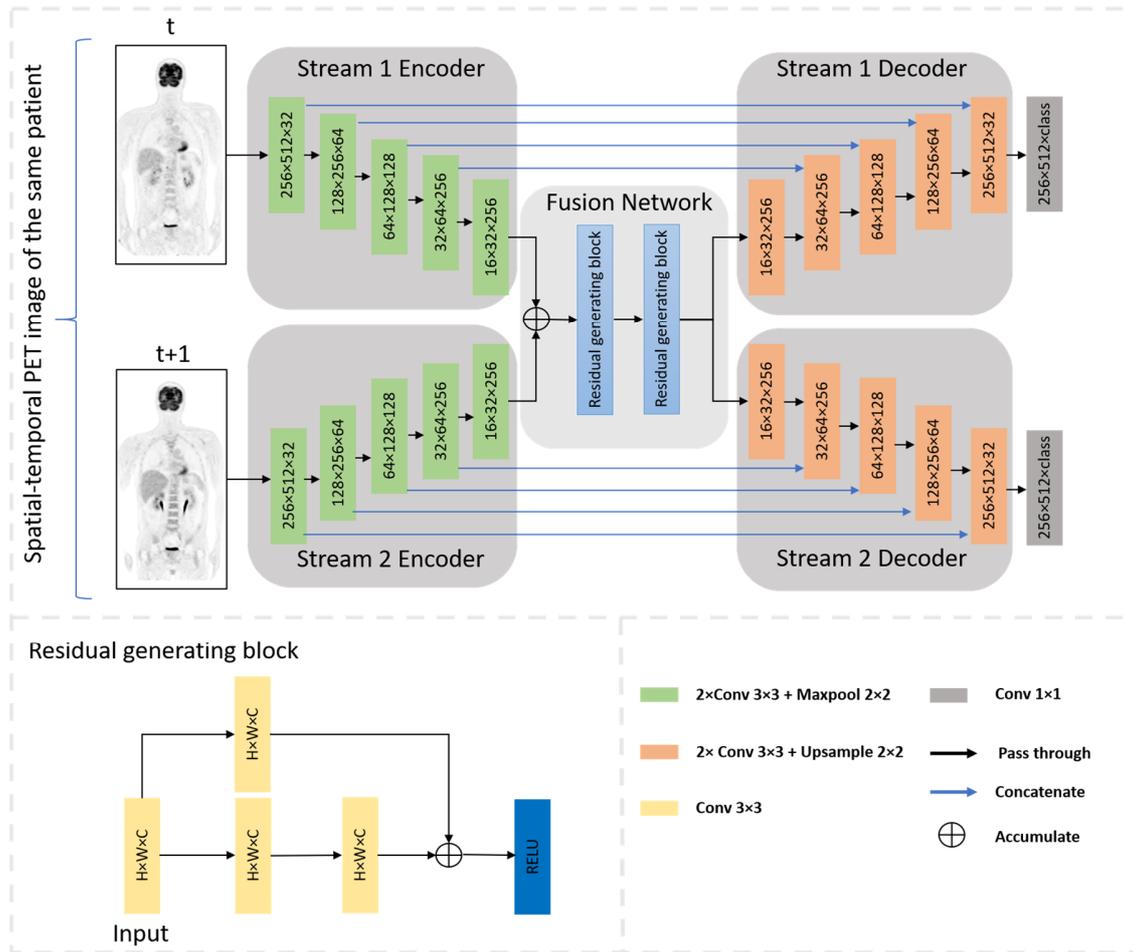

Figure 2. Our proposed ST-DSNN method.

Our ST-DSNN has five main components as shown in Fig. 2: two encoders, a fusion network, and two decoders. The encoders derive image features that can represent the target sFEPU at different time-points. The fusion blocks accumulate image features derived from each encoder to enhance regions with consistent characteristics while suppressing the inconsistent regions. The decoders integrate the accumulated features with features from the encoder across multiple image scales to generate the segmentation results.

**Encoder:** Our encoder consists of stacked convolutional layers. As shown in Fig 2, the encoder has one input convolutional block and four downsampling blocks. The input convolutional block contains two 3×3 convolutional layers to derive features from the input PET images and each of the four downsampling blocks contains one 2×2 maxpooling layer and two 3×3 convolutional layers to downsample derived the feature maps. We followed the well-established approach of using batch normalization [22] and rectified linear unit (ReLU) [23] activations after each convolutional layer. The two encoders shared weights and biases to ensure the extracted feature maps are aligned with each other. This ensures that feature maps that represent the consistent features in two time-points will appear in the same channel in the output of each encoder.

**Fusion Network:** The fusion network comprises two elements: (1) feature accumulation – combining the feature maps of the two time-points extracted by the encoders; and (2) residual generating block – deriving image features and generating fused feature maps for the two decoders. The inputs to the fusion network are two feature maps $F_{tp1}$ and $F_{tp2}$ (each the output of one encoder) each of size $w \times h \times c$ with $w$ width, $h$ height, and $c$ channels. The feature accumulation will integrate $F_{tp1}$ and $F_{tp2}$ elementwise to form the fusion feature as follow:

$$F_f = F_{tp1} + F_{tp2} \qquad\qquad (1)$$

where $F_f$ are the fusion feature maps with the size $w \times h \times c$. The feature map representing structures that are consistent across both time-points will appear in the corresponding channel in $F_{tp1}$ and $F_{tp2}$. Therefore, the features representing these temporally-consistent structures will be enhanced in $F_f$ while other features will be suppressed. The fused features will then be used as the input to the two residual generating blocks. We used the shortcut proposed by He et al. [24] but add a $3 \times 3$ convolutional in the shortcut. The purpose of the residual generating blocks is to further derive consistent features from the fusion features.

**Decoder:** Our decoder comprises four blocks; each block has one transpose convolutional layer and two convolutional layers. The transpose convolutional layer doubles the width and height of the input feature maps. The doubled feature maps were merged with the feature maps with the skip connection from the encoder and were then input into two $3 \times 3$ convolutional layers to avoid overfitting and to include features from the specific time-points to the decoder. After the last block of the decoder, the output feature map has the same width and height as the input PET image, but with 32 channels. We then used a $1 \times 1$ convolutional layer to project the feature maps to the number of sFEPU classes. Finally, we used the softmax function to transform this into a probability map representing the likelihood of individual pixels belonging to each sFEPU class. The weight sharing scheme was applied for the two decoders, in a similar configuration to the encoders.

### 2.4    Implementation Details

We implemented the ST-DSNN with Pytorch library [25]. The initial weights and biases were set via Xavier initialization [26]. We trained our ST-DSNN using the Adam optimizer [27] with a cross-entropy loss function. We set the training parameters empirically: the batch size was 6, the learning rate was $5 \times 10^{-5}$, weight decay was $1 \times 10^{-5}$ and the step size was 50. We applied random cropping as data augmentation, to improve the generalizability and to avoid overfitting. Our ST-DSSN was trained for 200 epochs on a workstation equipped with an 11GB NVIDIA GeForce 1080 Ti GPU.

### 2.3  Experimental Setup

We compared our ST-DSNN with: (1) U-Net [28] – consisting of an encoder and a decoder with skip connections; (2) VGG-FCN8s [29, 30] – the use of convolution kernel to extract image features of the input image and output the segmentation prediction (pixel to pixel) in an end-to-end manner. Different from U-Net, VGG-FCN8s used VGGNet as the network backbone; (3) RefineNet [31] – a multi-path refinement network that uses the information available along the down-sampling process and long-range residual connections to enable high-resolution image segmentation; (4) Combo-loss [18] – a 3D convolutional neural network (CNN) with a combination of two loss functions (dice and cross-entropy) as the loss function and, (5) ConvLSTM [21] – a convolutional LSTM for tumor growth prediction using sequential CT images. All comparison methods segmented 2D PET images, while only ConvLSTM used sequential information. All methods were trained and evaluated with the same five-fold cross-validation approach. For each-fold of the cross-validation, 8 patients' images (between 16 to 18 images per training fold) were used as the training, and 2 different patients' images (between 4 to 6 images per test fold) were used for testing.

We conducted experiments to assess the performance of different inputs of the ST-DSNN and to measure the use of spatio-temporal information. We compared our ST-DSNN with: (a) the same PET images in both encoders to measure the DSNN performance on the single time-point, (b) random unpaired images for both encoders when there is no consistency between two input images (e.g., co-segmenting PET images of the different patient in different time-points), and (c) the segmentation performance with different number of sequential PET images per patient.

### 2.4 Evaluation Metrics

The evaluation metrics we used were the Sørensen–Dice coefficient (DSC), Jaccard index (Jaccard), and positive position value (PPV). DSC and Jaccard evaluate the overlap ratio of the ground truth annotation and the predicted segmentation results, which is defined as:

$$DSC = \frac{2TP}{2TP + FP + FN} \tag{3}$$

$$Jaccard = \frac{TP}{TP + FP + FN} \tag{4}$$

Where $TP, FP,$ and $FN$ represent the number of true positive, false positive, and false negative pixels, respectively. PPV was used to measure the ratios of $TPs$ and $FPs$ was calculated as:

$$PPV = \frac{TP}{TP + FP}. \tag{5}$$

We used the $p$-value computed after a two-sample $t$-test to evaluate the significance of the results.

## 3. RESULTS

The comparison of sFEPU segmentation performance across all methods is presented in Table 2. ST-DSNN achieved the best overall performance across all three metrics. It outperformed the second-best method of U-Net by a margin of >1.8% in DSC, >2.2% in Jaccard, and >1.9% in PPV. Examples of the segmentation results are shown in Fig. 3.

Table 2. Segmentation results with comparison methods. Bold represents the best results. *represents methods using sequential images.

| | | Brain | Bladder | Heart | R.kidney | L.kidney | Mean |
|---|---|---|---|---|---|---|---|
| DSC | VGG-FCN8s | 94.80% | 93.17% | 93.23% | 74.85% | 80.47% | 87.30% |
| | U-Net | **94.85%** | **93.85%** | **93.40%** | 76.11% | 82.79% | 88.20% |
| | RefineNet | 91.32% | 79.49% | 81.64% | 48.77% | 46.67% | 69.58% |
| | Combo | 86.42% | 79.94% | 80.26% | 59.87% | 50.40% | 71.38% |
| | *ConvLSTM | 94.81% | 93.68% | 88.02% | 74.26% | 76.76% | 85.50% |
| | *ST-DSNN | 94.18% | 92.81% | 93.09% | **83.66%** | **86.40%** | **90.03%** |
| Jaccard | VGG-FCN8s | **90.60%** | 88.65% | **88.38%** | 63.68% | 69.76% | 80.22% |
| | U-Net | 90.47% | **90.42%** | 87.48% | 68.27% | 71.07% | 81.54% |
| | RefineNet | 86.29% | 73.84% | 74.07% | 39.20% | 36.96% | 62.07% |
| | Combo | 78.06% | 70.91% | 70.14% | 46.74% | 37.04% | 60.58% |
| | *ConvLSTM | 90.51% | 89.64% | 81.36% | 63.33% | 66.60% | 78.29% |
| | *ST-DSNN | 89.64% | 88.39% | 87.85% | **74.91%** | **78.06%** | **83.77%** |
| PPV | VGG-FCN8s | 94.34% | 94.67% | 93.41% | 82.90% | **88.17%** | 90.70% |
| | U-Net | 95.22% | **95.83%** | 86.02% | 79.82% | 89.61% |
| | RefineNet | 88.54% | 83.64% | 77.70% | 58.42% | 61.79% | 74.02% |
| | Combo | 89.26% | 86.86% | 84.03% | 68.56% | 63.92% | 78.53% |
| | *ConvLSTM | 94.11% | 93.88% | 87.58% | 78.60% | 73.92% | 85.62% |
| | *ST-DSNN | **95.73%** | 95.70% | **95.36%** | **89.19%** | 87.06% | **92.61%** |

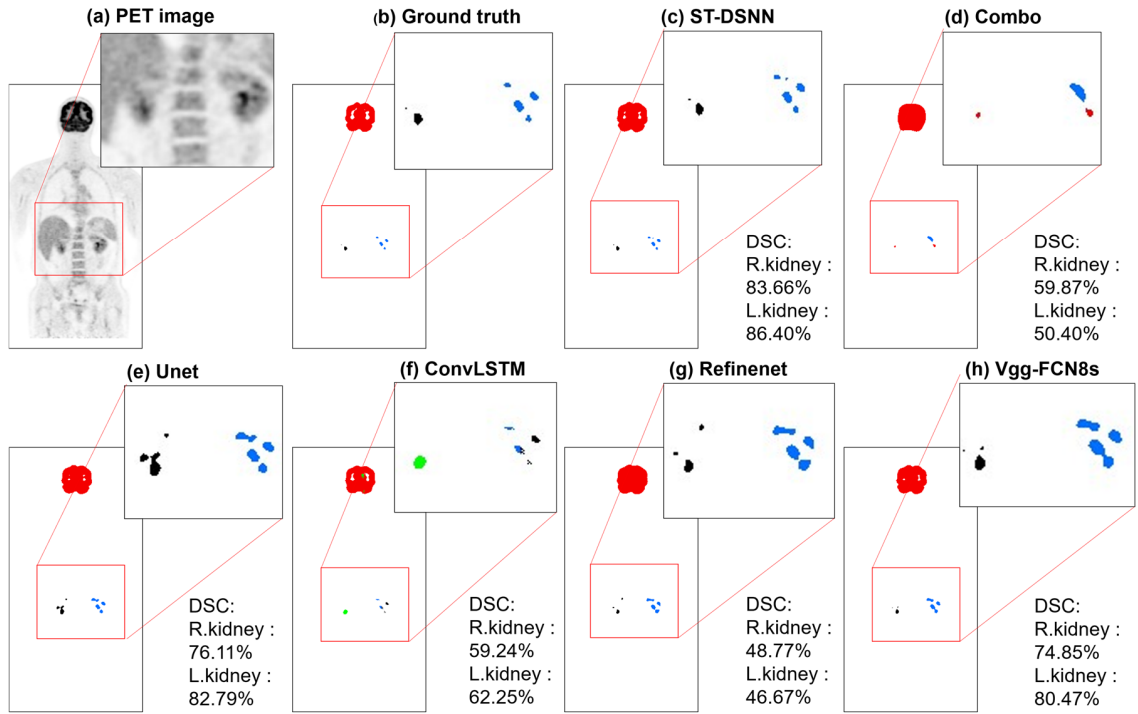

Figure 3. Segmentation results. (a) PET image (b) ground truth; (c-h) segmentation results derived from ST-DSNN, Combo, U-Net, ConvLSTM, RefineNet, and VGG-FCN8s methods.

The DSC results comparing the use of different input streams with ST-DSNN are shown in Table 3. The use of sequential images from adjacent time-points (default pairing) is superior to the use of the same images in each stream and the use of unpaired images. There is an improvement of 0.9% in the mean DSC and the improvement is statistically significant ($p < 0.01$). The 'same image' pair is a demonstration that our ST-DSNN can be used when only a single time-point image is available.

Table 3. Segmentation results of ST-DSNN with different input streams.

| DSC | Brain | Bladder | Heart | R.kidney | L.kidney | Mean |
|---|---|---|---|---|---|---|
| Same image | 95.19% | **94.83%** | 92.50% | 79.95% | 82.43% | 88.98% |
| Unpaired image | **95.76%** | 93.15% | **93.40%** | 80.77% | 82.51% | 89.12% |
| Sequential image (Default) | 94.18% | 92.81% | 93.09% | **83.66%** | **86.40%** | **90.03%** |

The segmentation performance with different number of sequential PET images is shown in Table 4. It shows that the proposed method has an improvement of 1.65% in mean DSC when more than 2 sequential PET images were used. In contrast, the comparison method ConvLSTM only has an improvement of 0.13% in DSC with additional sequential images.

Table 4. Segmentation results of DSC with ST-DSNN and ConvLSTM with additional sequential PET images.

|  | No. of Sequential Images | Brain | Bladder | heart | R.kidney | L.kidney | Mean |
|---|---|---|---|---|---|---|---|
| **ST-DSNN** | 2 | **94.92%** | 92.49% | 91.73% | 76.10% | **86.84%** | 88.42% |
|  | >2 | 93.02% | **93.98%** | **93.79%** | **84.84%** | 84.75% | **90.07%** |
| **ConvLSTM** | 2 | **94.88%** | 93.55% | **88.18%** | 73.99% | 76.71% | 85.47% |
|  | >2 | 94.65% | **93.98%** | 87.64% | **74.87%** | **76.86%** | **85.60%** |

## 4. DISCUSSION

Our main findings are that our ST-DSNN: (i) had more accurate sFEPU segmentation when compared to existing methods and, (ii) had consistent segmentation results across different sFEPU structures.

We attribute the overall best performance of the proposed ST-DSNN to the accumulation (fusion) of image features across sequential PET images (Table 2). The RefineNet had the poorest performance and we attribute this to the RefineNet using a linear interpolation process on the decoder side to upsample the derived image features and then generate the sFEPU segmentation results. The linear interpolation process was unsupervised and was unable to recover the information lost in the downsampling process. The better performance of the Combo [18] was because of the deconvolutional layers, on the decoder side, and the dual loss function. Specifically, the deconvolutional layers used learnable parameters to derive the segmentation results, which enabled a segmentation that was based on more meaningful features. In addition, Combo used both Dice loss and cross-entropy loss (dual loss function), which allowed consistent segmentation results compared with RefineNet which used a single loss function. The improvement of ConvLSTM to the Combo is expected. The ConvLSTM leverages the input gate to augment the learnable image features with sequential PET images and thus has more features to learn when compared with methods used for single time-points. The output gate, which is the compartment to decide which information to deliver, was designed to capture the subtle changes within different time-points [21]. Consequently, ConvLSTM failed to segment the sFEPU structures when there were large physiologic changes e.g., kidneys when there was asymmetric excretion. Our ST-DSNN, when compared to the second best-performing method of U-Net and third best-performing FCN-VGG8s, performed better in segmenting the kidneys. Our approach had a slightly poorer performance in segmenting the brain, bladder and the heart; these structures are relatively easy to segment and are the dominant structures (more training samples e.g., pixels) in our sFEPU. Consequently, both U-Net and FCN-VGG8s tend to overfit to these dominant structures and failed to segment the kidneys which have fewer training samples. In contrast, given that our approach leverages sequential images, it accumulates image features of different sFEPUs across multiple scans and so ensures that the training image features are balanced across dominant / non-dominant structures to give consistent segmentation results.

The segmentation results derived with unpaired images have better performance in overall DSC when compared to segmentation derived from the same images. This is an expected finding because unpaired images add more image feature variations to the training dataset. The accumulated image features of unpaired images, however, may have image features derived from two different patients, which are then not relevant to the sFEPU to be segmented. Our approach meanwhile uses the image sequence from the same patient which resulted in better performance.

The segmentation results for both the ST-DSNN and the ConvLSTM improved with more image sequences, with ST-DSNN benefitting more from the addition. We attribute this better performance to the fact that additional sequential PET images bring additional image features of the same patient that can be accumulated, which are then used for more representative sFEPU image features. In contrast, ConvLSTM was designed to identify changes in sequential data, which thereby has limited capability to leverage the additional image features bought by the additional sequential PET images.

Our focus in the current study was to investigate the use of spatio-temporal features derived CNNs for segmenting the sequential PET images. Whilst our approach has improved performance compared to existing methods and we have not generalized our results to other tumor types or where other imaging modalities are employed. In future work, we intend to test our approach in non-small cell lung cancer and the soft-tissue sarcomas, using PET, and also other imaging modalities such as CT and magnetic resonance (MR).

## 5. Conclusions

We proposed a spatio-temporal dual-stream neural network (ST-DSNN) method to segment sequential PET images. Our method achieved better segmentation results through accumulating consistent spatio-temporal features from sequential PET scans and had better segmentation accuracy when compared to the state-of-the-art methods.

## Acknowledgment

This work was supported in part by Australian Research Council (ARC) grants (IC170100022 and DP170104304).

## References

1. Hingorjo, M.R. and S. Syed, *Presentation, staging and diagnosis of lymphoma: a clinical perspective.* Journal of Ayub Medical College, Abbottabad : JAMC, 2008. **20**(4): p. 100-103.

2. Berriolo-Riedinger, A., et al., *Role of FDG PET-CT in the treatment management of Hodgkin lymphoma.* Cancer/Radiotherapie, 2018. **22**(5): p. 393-400.

3. Sher, A.C., et al., *Assessment of Sequential PET/MRI in Comparison With PET/CT of Pediatric Lymphoma: A Prospective Study.* American Journal of Roentgenology, 2016. **206**(3): p. 623-631.

4. Shreve, P.D., Y. Anzai, and R.L. Wahl, *Pitfalls in Oncologic Diagnosis with FDG PET Imaging: Physiologic and Benign Variants.* RadioGraphics, 1999. **19**(1): p. 61-77.

5. La Fontaine, M., et al., *A secondary analysis of FDG spatio-temporal consistency in the randomized phase II PET-boost trial in stage II–III NSCLC.* Radiotherapy and Oncology, 2018. **127**(2): p. 259-266.

6. Vauclin, S., et al., *Development of a generic thresholding algorithm for the delineation of 18FDG-PET-positive tissue: Application to the comparison of three thresholding models.* Physics in Medicine and Biology, 2009. **54**(22): p. 6901-6916.

7.      Drever, L., et al., *Comparison of three image segmentation techniques for target volume delineation in positron emission tomography.* Journal of Applied Clinical Medical Physics, 2007. **8**(2): p. 93-109.

8.      Ju, W., et al., *Random walk and graph cut for co-segmentation of lung tumor on PET-CT images.* IEEE Transactions on Image Processing, 2015. **24**(12): p. 5854-5867.

9.      Gubbi, J., et al., *Automatic tumour volume delineation in respiratory-gated PET images.* Journal of medical imaging and radiation oncology, 2011. **55**(1): p. 65-76.

10.     Aristophanous, M., et al., *A Gaussian mixture model for definition of lung tumor volumes in positron emission tomography.* Medical Physics, 2007. **34**(11): p. 4223-4235.

11.     Tajbakhsh, N., et al., *Convolutional Neural Networks for Medical Image Analysis: Full Training or Fine Tuning?* IEEE Transactions on Medical Imaging, 2016. **35**(5): p. 1299-1312.

12.     Zhang, W., et al., *Deep convolutional neural networks for multi-modality isointense infant brain image segmentation.* NeuroImage, 2015. **108**: p. 214-224.

13.     Zhao, L., et al., *Automatic Nasopharyngeal Carcinoma Segmentation Using Fully Convolutional Networks with Auxiliary Paths on Dual-Modality PET-CT Images.* Journal of Digital Imaging, 2019. **32**(3): p. 462-470.

14.     Zhong, Z., et al., *Simultaneous cosegmentation of tumors in PET-CT images using deep fully convolutional networks.* Medical Physics, 2019. **46**(2): p. 619-633.

15.     Zhao, X., et al., *Tumor co-segmentation in PET/CT using multi-modality fully convolutional neural network.* Physics in Medicine and Biology, 2019. **64**(1).

16.     Blanc-Durand, P., et al., *Automatic lesion detection and segmentation of18F-FET PET in gliomas: A full 3D U-Net convolutional neural network study.* PLoS ONE, 2018. **13**(4).

17.     Bi, L., et al., *Automatic detection and classification of regions of FDG uptake in whole-body PET-CT lymphoma studies.* Computerized Medical Imaging and Graphics, 2017. **60**: p. 3-10.

18.     Taghanaki, S.A., et al., *Combo loss: Handling input and output imbalance in multi-organ segmentation.* Computerized Medical Imaging and Graphics, 2019. **75**: p. 24-33.

19.     Hu, H., et al. *Lymphoma Segmentation in PET Images Based on Multi-view and Conv3D Fusion Strategy.* in *Proceedings - International Symposium on Biomedical Imaging.* 2020.

20.     Debus, A. and E. Ferrante, *Left Ventricle Quantification Through Spatio-Temporal CNNs*, in *Lecture Notes in Computer Science (including subseries Lecture Notes in Artificial Intelligence and Lecture Notes in Bioinformatics)*. 2019. p. 466-475.

21.     Zhang, L., et al., *Spatio-Temporal Convolutional LSTMs for Tumor Growth Prediction by Learning 4D Longitudinal Patient Data.* IEEE Transactions on Medical Imaging, 2020. **39**(4): p. 1114-1126.

22.     Ioffe, S. and C. Szegedy, *Batch normalization: Accelerating deep network training by reducing internal covariate shift.* arXiv preprint arXiv:1502.03167, 2015.

23.     Nair, V. and G.E. Hinton. *Rectified linear units improve restricted boltzmann machines.* in *ICML.* 2010.

24.     He, K., et al., *Deep residual learning for image recognition.* 2016, IEEE Computer Society. p. 770-778.

25.     Paszke, A., et al. *Pytorch: An imperative style, high-performance deep learning library.* in *Advances in neural information processing systems.* 2019.

26.     Glorot, X. and Y. Bengio, *Understanding the difficulty of training deep feedforward neural networks.* Journal of Machine Learning Research, 2010. **9**: p. 249-256.

27.     Kingma, D.P. and J. Ba, *Adam: A method for stochastic optimization.* arXiv preprint arXiv:1412.6980, 2014.

28.     Ronneberger, O., P. Fischer, and T. Brox, *U-net: Convolutional networks for biomedical image segmentation*, in *Lecture Notes in Computer Science (including subseries Lecture Notes in Artificial Intelligence and Lecture Notes in Bioinformatics)*. 2015. p. 234-241.

29.     Long, J., E. Shelhamer, and T. Darrell, *Fully convolutional networks for semantic segmentation.* 2015, IEEE Computer Society. p. 3431-3440.

30.     Simonyan, K. and A. Zisserman, *Very deep convolutional networks for large-scale image recognition.* arXiv preprint arXiv:1409.1556, 2014.

31.     Nekrasov, V., C. Shen, and I. Reid. *Light-weight refinenet for real-time semantic segmentation.* in *British Machine Vision Conference 2018, BMVC 2018.* 2019.